\documentclass{emulateapj}

\usepackage{amssymb,amsmath} 

\newcommand\beq{\begin{equation}}
\newcommand\eeq{\end{equation}}
\newcommand\beqar{\begin{eqnarray}}
\newcommand\eeqar{\end{eqnarray}}

\begin{document}

\title{A SEARCH FOR CO-EVOLVING ION AND NEUTRAL GAS SPECIES IN PRESTELLAR MOLECULAR CLOUD CORES}

\author{Konstantinos Tassis\altaffilmark{1}, 
Talayeh Hezareh\altaffilmark{1},
\& Karen Willacy\altaffilmark{2} }

\altaffiltext{1}{Max-Planck Institut f\"ur Radioastronomie, 53121 Bonn, Germany}
\altaffiltext{2}{Jet Propulsion Laboratory, California Institute of Technology, Pasadena, CA 91109, USA}

\begin{abstract}

Comparison of linewidths of spectral line profiles of ions and neutral molecules have been recently used to estimate the strength of the magnetic field in turbulent star-forming regions. However, the ion (HCO$^+$) and neutral (HCN) species used in such studies may not be necessarily co-evolving at every scale and density and may thus not trace the same regions. Here, we use coupled chemical/dynamical models of evolving prestellar molecular cloud cores including non-equilibrium chemistry, with and without magnetic fields, to study the spatial distribution of HCO$^+$ and HCN, which have been used in observations of spectral linewidth differences to date. In addition, we seek new ion-neutral pairs that are good candidates for such observations because they have similar evolution and are approximately co-spatial in our models. We identify three such good candidate pairs: HCO$^+$/NO, HCO$^+$/CO, and NO$^+$/NO. 
\end{abstract}


\section{Introduction}

Observations of large velocity dispersions in the line profiles of molecules (line widths wider than expected from thermal motions) in interstellar molecular clouds is a signature of localized supersonic turbulence, while observations of the Zeeman effect (e.g., \citealp{Crutcheretal1999}) and the polarization of dust thermal emission (e.g., \citealp{Hildebrandetal2009}) indicate that these clouds are permeated by magnetic fields. Thus interstellar molecular clouds need to be studied in the framework of magnetohydrodynamics. 

For typical molecular cloud conditions (e.g., $B \sim 10 \mu G$, density $\sim 10^3 {\rm cm^{-3}}$, and ionized fraction $\lesssim 10^{-6}$) magnetic forces, which are experienced directly by ionized particles, are transmitted to the bulk of the neutral gas molecules via frequent collisions. However the coupling of the magnetic field with the gas is not perfect. Motions of the bulk neutral gas with frequencies higher than the typical ion-neutral collision frequency can not be transfered to the ions.
This relative motion between charged and neutral particles, known as ambipolar drift \citep{MS56}, results in the dissipation of hydromagnetic waves in small scales \citep{Brag65,KP69,ZJ83} and can cause differences between the velocity dispersion spectra of the ions and the neutrals, as was recognized by \citet{Houdeetal2000a, Houdeetal2000b} and explored further by \citet{LiHoude08}.
Indeed, \citet{Mouschovias2011},  in a comprehensive study of magnetohydrodynamic waves in weakly-ionized gases, showed that  most hydromagnetic waves with wavelengths below a critical value cannot propagate in the neutrals because they are damped rapidly by ambipolar diffusion, but certain long-lived modes in the neutrals exist at all wavelengths; in contrast, waves in the ions damp very rapidly at short wavelengths. 

Observationally, the narrower line profiles of ions compared to that of co-existing neutral species was first noticed and investigated by \citet{Houdeetal2000a,Houdeetal2000b}, who argued that the reason of this ion line narrowing effect was due to the presence of magnetic fields in turbulent molecular clouds that are on average not aligned with the local flows. To test this analysis, \citet{Lai2003} observed DR21(OH) at a high spatial resolution in H$^{13}$CO$^+$ and H$^{13}$CN  and confirmed that the ion-neutral pair is co-existent within their observing scale ($6^{\prime\prime}$) and that the ion line widths are indeed narrower than the neutral line widths. \citet{LiHoude08} re-visited this magnetically induced ion line narrowing effect. They mapped the M17 star forming region in ${\rm HCO^+}$ $(J = 4 \rightarrow 3)$ and HCN $(J = 4 \rightarrow 3)$ and fitted the lower envelopes of the velocity dispersion spectra of the observed line profiles of the ion and neutral pair to a Kolmogorov-type power-law. They determined the ambipolar diffusion length scale using the fitting parameters,  and used it to calculate the strength of the plane-of-sky component of the magnetic field. \citet{Hezareh2010} further tested the \citet{LiHoude08}  technique by mapping DR21(OH) in the optically thin ${\rm H^{13}CO^+}$ and ${\rm H^{13}CN}$ $(J = 4 \rightarrow 3)$ lines and obtained the turbulent ambipolar diffusion length and magnetic field strength in that source. \citet{Hezareh2012} used interferometry maps of the ground transition of the latter pair of molecules in a few massive dense cores in the Cygnus X region to verify that the method is reproducible for both single-dish and interferometry data. 

One concern in interpreting these observations is that chemical differentiation might also be responsible for differences in the spectral appearance of different molecular species: species with different spatial distributions would also exhibit different velocity profiles. 
Although HCN and HCO$^+$ have been shown to be approximately co-existent in the observed molecular clouds on the scales probed \citep{Houdeetal2000a,Houdeetal2000b,Lai2003,Houdeetal2004}, their chemical co-evolution  has yet to be examined from a theoretical perspective. Here, we use both magnetic and non-magnetic dynamical models of core collapse coupled with, and self-consistently following, non-equilibrium chemistry both in the gas phase and on grains (\citealp{Tassisetal2012a}, hereafter paper I) to examine whether the typically observed neutrals and ions are indeed theoretically expected to be similarly distributed in a star-forming core. We also repeat this exercise for other frequently observed neutral and ionized molecules, and we propose neutral-ion pairs that, according to our models, best maintain a co-spatial distribution.

\section{Models}\label{mod}

\begin{figure}
%\epsscale{1.2}
\plotone{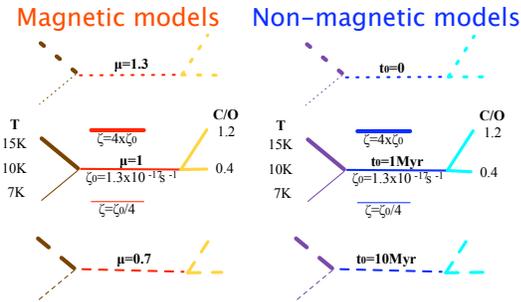}
\caption{\label{ModVis} 
Line types and colors used to denote each of the models studied. Solid normal-thickness red line: ``reference'' magnetic model; solid normal-thickness blue line: ``reference'' non-magnetic model. Dotted lines: ``fast'' models; dashed lines: ``slow'' models. Brown/purple lines: magnetic/nonmagnetic models with temperatures differing from the ``reference'' models. Orange/cyan shaded areas: variation in C/O ratio. Thin/thick solid red/blue lines: lower/higher cosmic-ray ionization rate magnetic/non-magnetic models (see text for details). }
\end{figure}

We are interested in the co-spatiality of various observable neutral-ion pairs as a function of density in a star-forming region. Since the behavior of the abundance of many molecules with density is sensitive to various model parameters,  we use all Paper I models to assess the behavior of the abundance {\em ratios} of neutral-ion pairs with density. The best candidate molecule pairs for studies of systematic differences of the linewidths between neutrals and ions will be those with abundance ratios that vary weakly with density for any set of model parameters. 

The physics of the chemodynamical models we use and its effect on molecular abundances are exhaustively discussed in Paper I. Here, we briefly review the parameters we have varied for each class of models: temperature, C/O ratio, cosmic ray ionization rate, and a parameter 
controlling the time available for chemical evolution. The latter is the mass-to-magnetic-flux ratio for magnetic models, and the collapse delay time for non-magnetic models (an initial time period during which chemistry evolves but the core does not 
evolve dynamically, representing an early stage of support due to turbulence which later decays). 

The initial parameter values for our ``reference'' magnetic model are a  mass-to-magnetic-flux ratio equal to the critical value for collapse \citep{MouschoviasSpitzer76}, a temperature of 10 K, a C/O ratio of 0.4, and the canonical value for the cosmic ray ionization rate $\zeta = 1.3 \times 10^{-17} {\rm \, s^{-1}}$. The relevant parameter values of our ``reference'' non-magnetic model are a collapse delay time (hereafter ``delay'') of 1 Myr, with the other parameters being identical to the ``reference'' magnetic model. 
 
 For magnetic models we vary the initial mass-to-magnetic-flux ratio, examining two additional values: 1.3 times the critical value 
(a faster-evolving, magnetically supercritical model), and 0.7 of the critical value (a slower, 
magnetically subcritical model). For non-magnetic models we examine two additional values of delay: zero, and 10 Myr. 

For each of these six dynamical models,  the carbon-to-oxygen ratio is varied from its ``reference'' value by keeping the abundance of C constant and changing that of O. Additional values of C/O ratio we examined are 1 and 1.2. This way, we study a ``matrix'' of 18 combinations (9 magnetic and 9 non-magnetic) of available times for chemical evolution and C/O ratios.

To test the effect of temperature ($T$), we have varied each of the six basic dynamical models by changing $T$ by a factor of  $\sim 1.5$ from its reference value of 10 K and examined models with $T = 7$ K and $T = 15$ K. This results in 12 additional models (6 magnetic and 6 non-magnetic).

To test the cosmic ray ionization variation, we studied four additional models (two magnetic and two non-magnetic), which have the ``reference'' value for the temperature, C/O ratio, and mass-to-flux ratio or delay (for magnetic and non-magnetic models respectively), but for which $\zeta$ is varied by a factor of four above ($\zeta = 5.2 \times 10^{-17}$ $s^{-1}$) and below ($\zeta = 3.3 \times 10^{-18}$ $s^{-1}$) its ``reference'' value, covering the range of observational estimates (e.g., \citet{McCall2003,Hezareh2008}).

Figure \ref{ModVis} shows a visual representation of these suite of models and is a quick reference guide for the line colors and types we have used to depict each model.

\section{Results}\label{res}

\subsection{\rm{HCO$^+$/HCN}}

\begin{figure*}
%\epsscale{0.7}
%\plotone{plotcentabund_HCOpHCNrat.eps}
\plotone{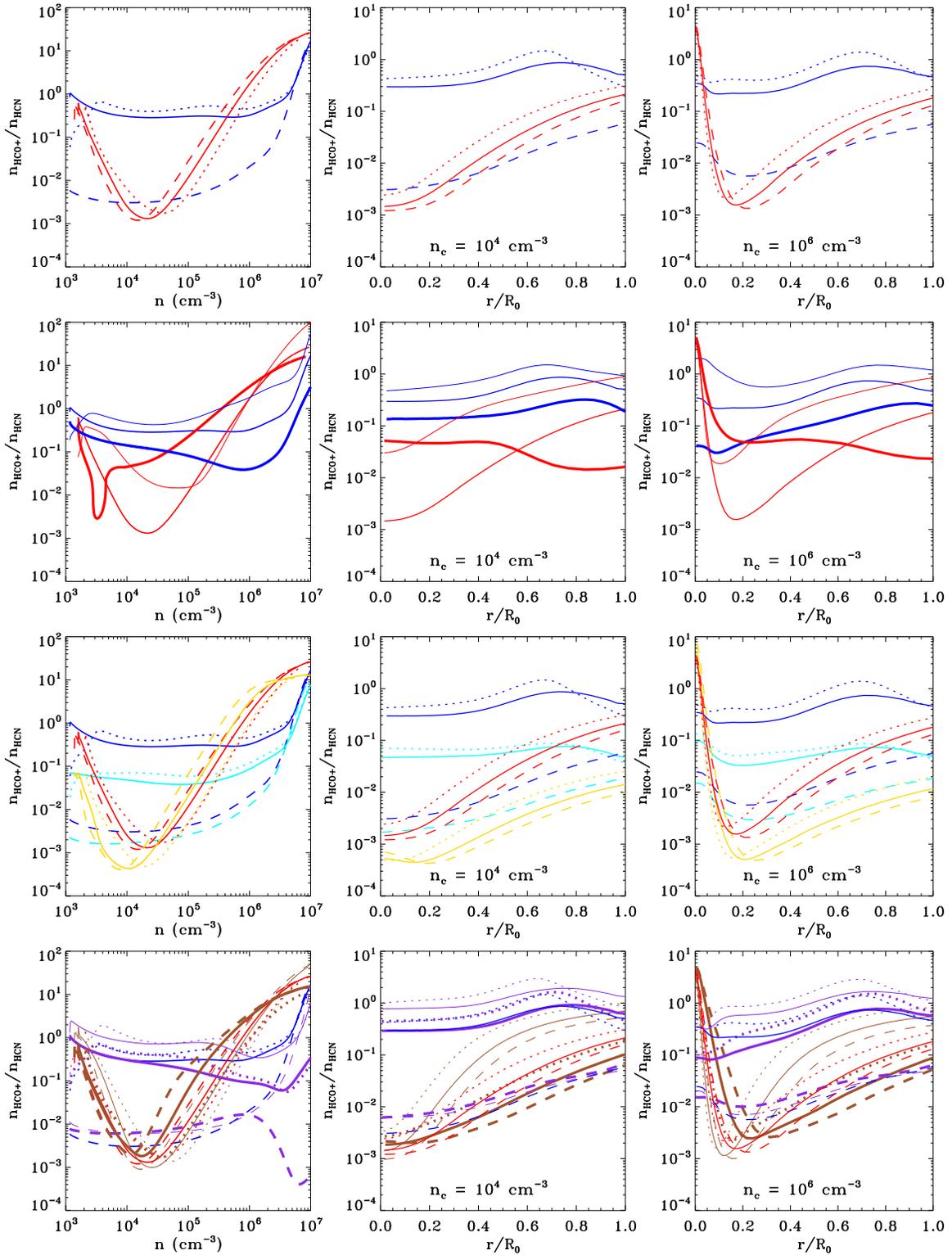}
\caption{\label{obsratios} 
Left column: Evolution of central abundance ratio of HCO$^+$ and HCN as a function of central number density. Middle and right columns: radial profiles of the HCO$^+$ / HCN abundance ratio, for two different evolutionary stages as quantified by the central number density. Line types and colors are as described in Fig.~\ref{ModVis}. Top row: Different dynamical models for fiducial values of the C/O ratio, CR ionization rate, and temperature. Seond row: effect of the CR ionization rate for the fiducial magnetic and non-magnetic models. Third row: effect of the C/O ratio for each dynamical model, for fiducial values of the CR ionization rate and temperature. Bottom row: effect of temperature for each dynamical model for fiducial values of the CR ionization rate and C/O ratio. }
\end{figure*}

\begin{figure*}
%\plotone{plotcentabund_HCOpNOrat.eps}
\plotone{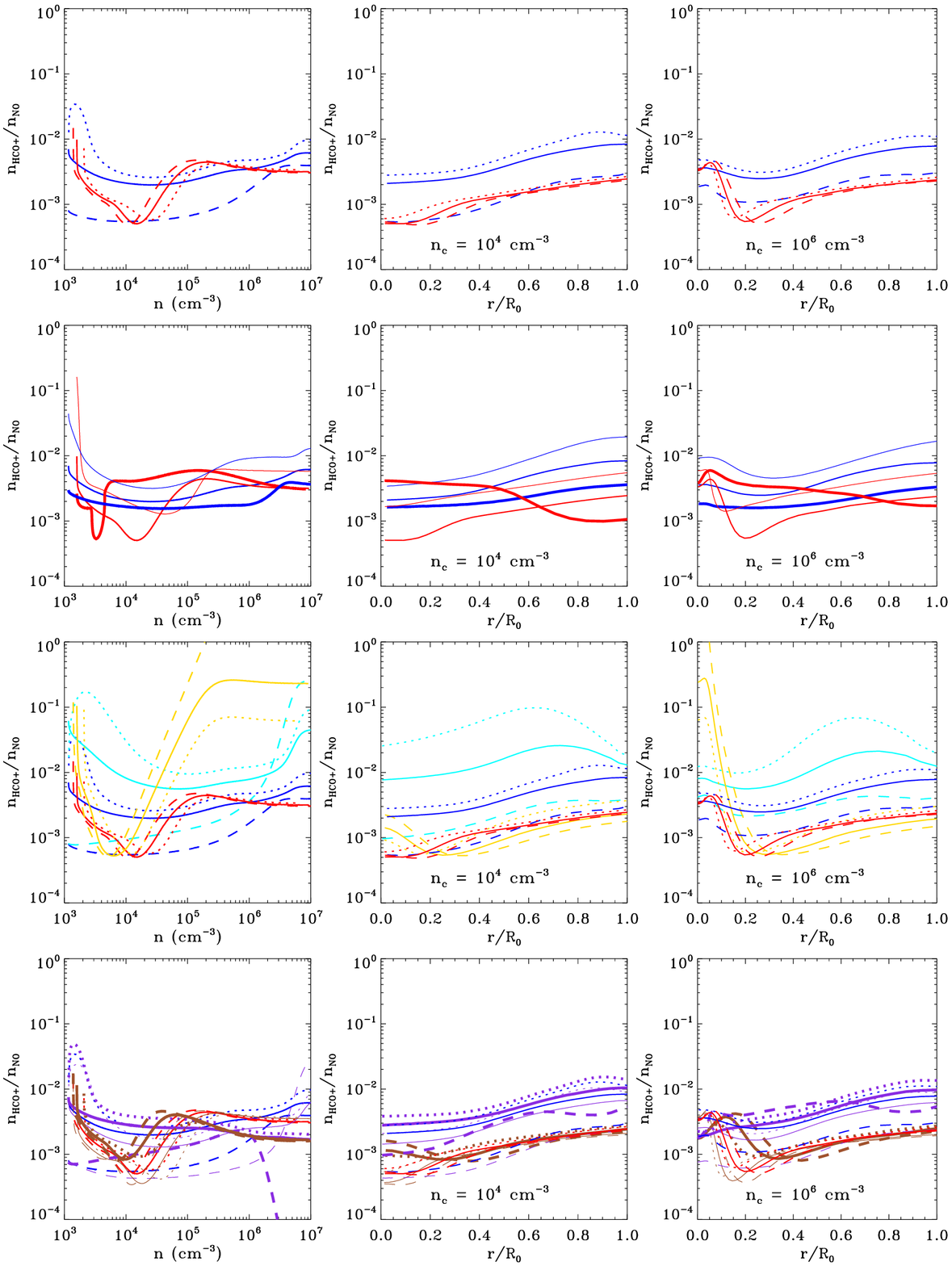}
\caption{\label{ratioNO} 
As in Fig.~\ref{obsratios} but for the HCO$^+$ / NO pair.}
\end{figure*}

\begin{figure*}
%\plotone{plotcentabund_HCOpCOrat.eps}
\plotone{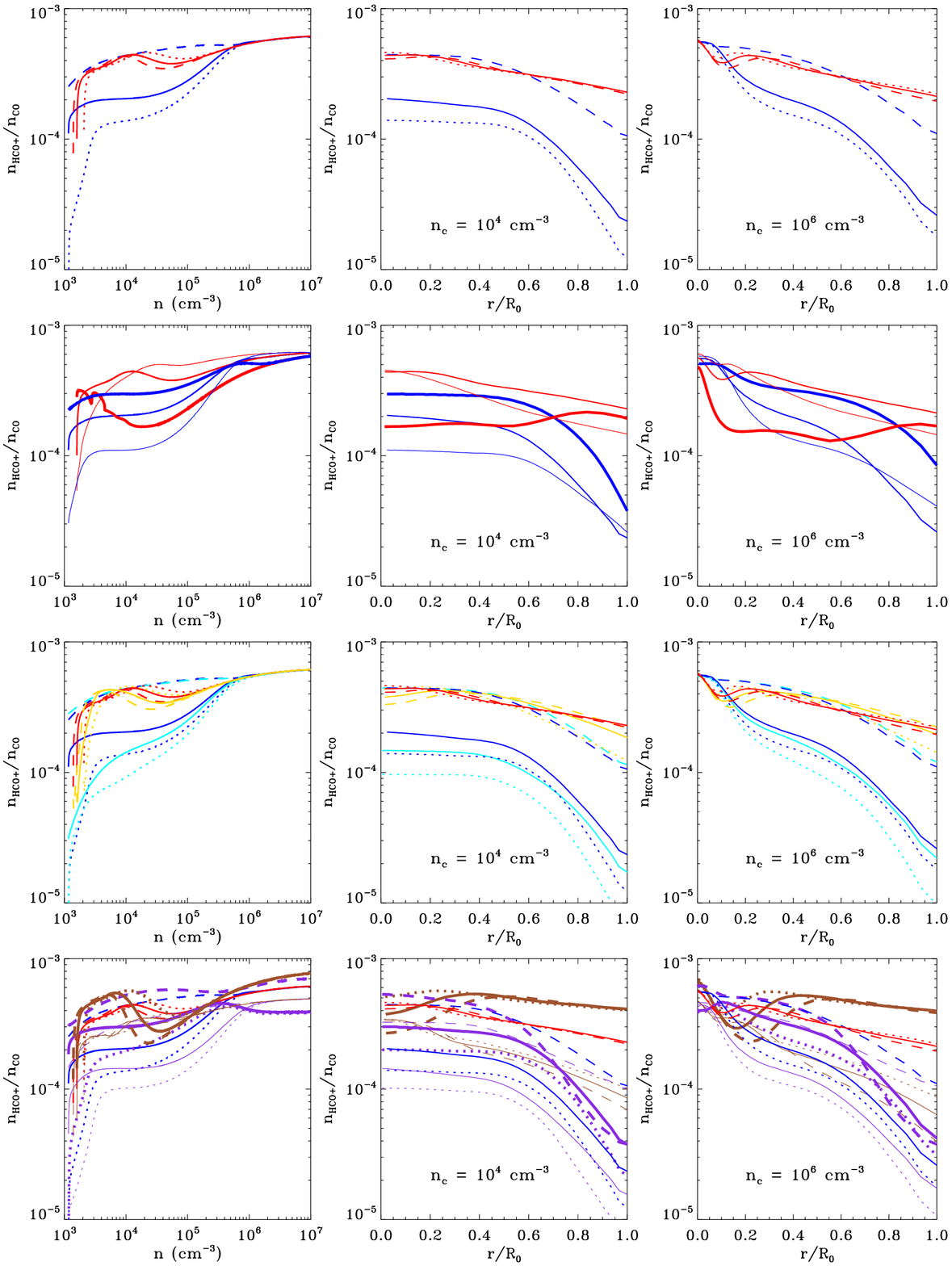}
\caption{\label{ratioCO} 
As in Fig.~\ref{obsratios} but for the HCO$^+$ / CO pair.}
\end{figure*}

\begin{figure*}
%\plotone{plotcentabund_NOpNOrat.eps}
\plotone{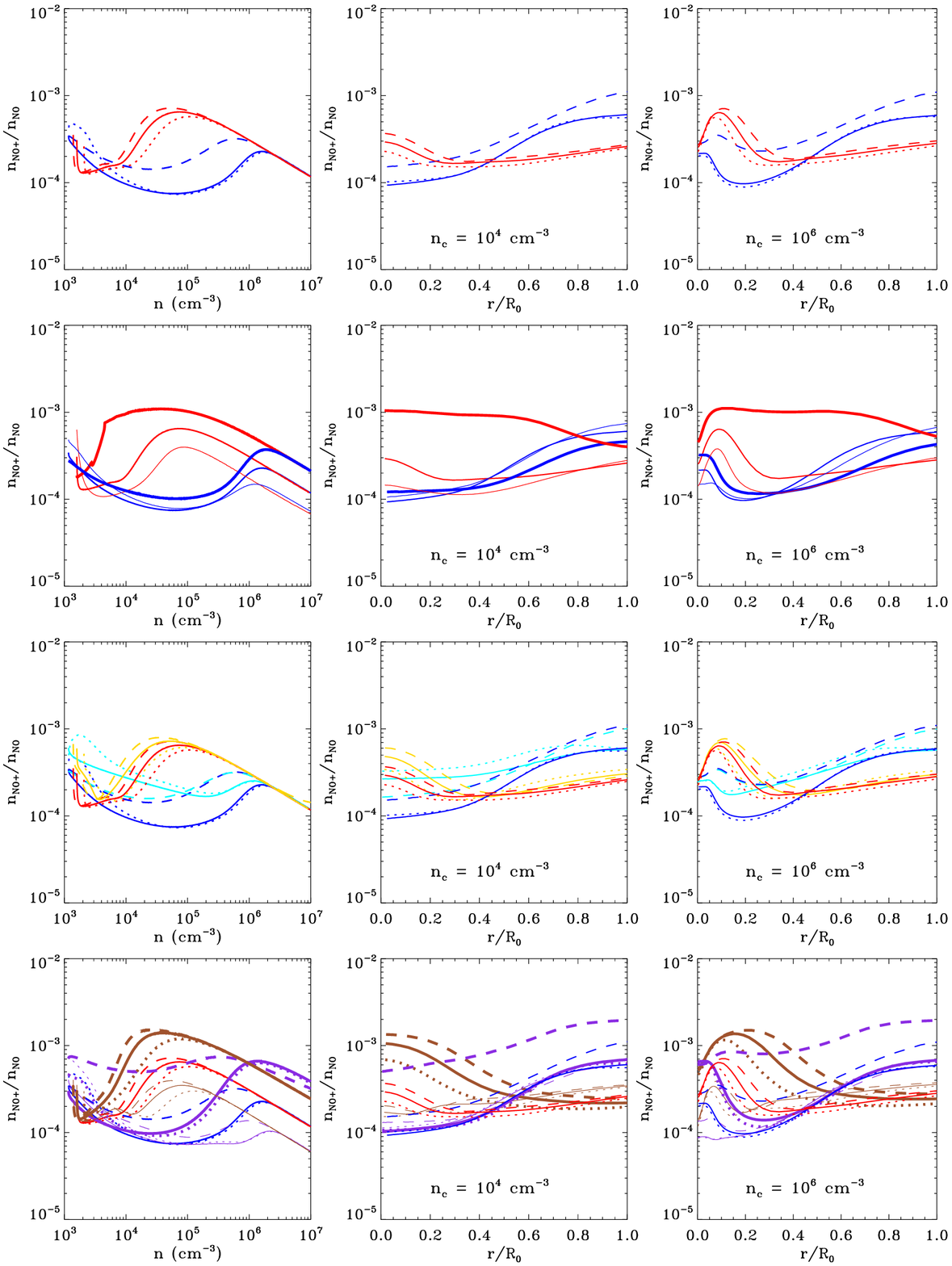}
\caption{\label{ratioNOp} 
As in Fig.~\ref{obsratios} but for the NO$^+$ / NO pair.}
\end{figure*}

Figure \ref{obsratios} shows the results of our aforementioned models for the evolution of the commonly observed ion-neutral pair HCO$^+$/HCN. We show the evolution of the central abundance ratio with central density (left column), and radial profiles of the abundance ratio at two different evolutionary stages (of central density $10^4$ and $10^6$ cm$^{-3}$, central and right columns, respectively).  The top row shows the effect of dynamics on the abundance ratio, while the second, third, and bottom rows demonstrate the effect of the CR ionization rate, C/O ratio, and temperature, respectively. 

As shown in the left column, the central abundance ratio, especially in the magnetic models, is not particularly stable with central density. The reason, as can be seen in Figure 5 of Paper I, is that although both HCO$^+$ and HCN deplete with increasing central density, in our magnetic models HCO$^+$ depletes initially faster  than HCN and therefore their ratio declines by almost three orders of magnitude by a central density of $10^4$ cm$^{-3}$. However, at higher central densities, the depletion rate of HCO$^+$ slows down while HCN keeps depleting at its initial rate. This effect is responsible for the increase of the abundance ratio at higher densities. In non-magnetic models, the dependence of the abundance ratio on central density is much less pronounced, as the behavior of each of HCO$^+$ and HCN with central density is qualitatively similar (Figure 5 of Paper I), with only a small rise in the abundance ratio above $10^6$ cm$^{-3}$ due to some flattening of the depletion rate of HCO$^+$ at the highest densities examined. 

This behavior is also reflected in the radial profiles of this abundance ratio, as seen in the middle and right columns of Figure \ref{obsratios}. The magnetic models and the slow non-magnetic models (blue, cyan, and purple dashed curves) exhibit a remarkable variation between the outer layers of the star-forming core where significant amounts of HCO$^+$ still exist, and the innermost and densest parts. The fast non-magnetic models, however, have a much milder dependence on radius. At a more advanced evolutionary stage (n$_c \sim 10^6$ cm$^{-3}$), the minimum seen in the evolution of the central abundance ratio with density in magnetic models also appears in the radial profile at intermediate scales. At scales comparable to the size of the core, the radial profiles of the HCO$^+$/HCN abundance ratio flatten out for all models.

The total variation of the abundance ratio across models, parameters, and densities is large (several orders of magnitude). As seen in the top row of Fig.~\ref{obsratios}, most of the total variation has its origin in the dynamics, with the CR ionization rate having the second most important contribution (second row). However, different models represent different possible clouds in nature, and for this reason the variation of the abundance ratio across models is not the relevant concern for the Houde et al. type of analysis; instead, the quantity of interest is the total variation of the ratio {\em within a single model}, across space and time. Still, as seen in for example in the left column of Fig.~\ref{obsratios}, the maximum possible variation within a single model can be as large as several orders of magnitude. 

Observationally, the intensity maps of the HCO$^+$/HCN pair show highest degree of correlation at large length scales, about $0.5-1.0$ pc \citep{Lo2009}, which correspond to scales comparable with or larger than the entire core in our models ($r/R_0 \gtrsim 1$ in Figure \ref{obsratios}). However, this correlation can be reduced at smaller scales of about $0.1$ pc, i.e., resolutions achieved by interferometric observations \citep{Csengeri2011}.

HCO$^+$ and HCN follow independent chemical networks and thus have no reason to be co-evolving. The two molecules evolve differently and cannot be considered co-spatial in most of our models. HCO$^+$ is more abundant at lower densities and larger spatial scales.  Furthermore, the critical densities of  the commonly observed rotational transitions of  HCO$^+$ are lower than these of HCN. Consequently, HCO$^+$  would in principle be expected to trace larger spatial scales, regions with higher levels of turbulence, which would lead to wider linewidths for HCO$^+$ rather than HCN. However, this is contradictory to observations. \citet{Houdeetal2000a} observed the $J=3\rightarrow2$ and $J=4\rightarrow3$ transitions of HCO$^+$ and HCN in several molecular clouds where they found these species co-existent, with the ion lines indeed more opaque but {\em narrower} than the neutral lines. This can be explained by the effect of the magnetic field on the dynamics of molecular ions through the cyclotron interaction, which can in turn be interpreted as a signature of ambipolar diffusion \citep{Houdeetal2004} and be used to estimate the strength of the magnetic field. Indeed, it was later shown that clouds with stronger magnetic fields (projected in the plane of the sky) show larger differences between the ion and neutral spectral linewidths \citep{Lietal2010,Hezareh2012}. In contrast, chemical and radiative transfer effects act in the opposite direction  causing  the widths of neutral spectral lines to become narrower than those of ions, thus underestimating the strength of the magnetic field.

To make a more robust comparison of our models with existing observations we must perform radiative transfer calculations for the lines observed. We will return to this problem in the future. An observational correlation similar to HCO$^+$ and HCN is observed for the H$^{13}$CO$^+$ and H$^{13}$CN pair but since our chemical models do not incorporate the chemistry of the isotopes of carbon and other elements, we do not include these isotopologue pairs in our discussion. For the remainder of this paper, we identify additional ion and neutral species that are good candidates for being considered as co-existent pairs.

\subsection{\rm{HCO$^+$/NO}}

Figure \ref{ratioNO} shows the same pattern as in Figure~\ref{obsratios}, i.e., the evolution of central abundance ratio with central density in the left column, and radial profiles of the abundance ratio for central densities of $10^4$ and $10^6 {\rm \, cm^{-3}}$ in the middle and right columns, respectively, for the HCO$^+$ /  NO pair. The central abundance ratio of this pair is significantly more stable with central density within any single model than the HCO$^+$/HCN ratio, as HCO$^+$ and NO evolve in a similar manner with central density, with the exception of the high C/O ratio models (third row: yellow and cyan lines for the magnetic and non-magnetic models, respectively). In all other cases, the variation of the abundance ratio with central density within a single model is $\lesssim$ an order of magnitude over the central density range we examine. 

A similar behavior is reflected in the radial profiles of the HCO$^+$ /  NO  abundance ratio, with most models exhibiting little variation. The large variation in the HCO$^+$ / NO pair for the magnetic high C/O-ratio model (yellow curve) only affects very small scales, which are also associated with high densities. At these densities, however, the depletion of these molecules is significant (Fig. 5 in Paper I), so these regions are not expected to have a considerable contribution to the observed lines.

\subsection{\rm{HCO$^+$/CO}}

The plots for the HCO$^+$ / CO pair is shown in Figure ~\ref{ratioCO}. This pair exhibits very small variations of the central abundance ratio with central density for all but the fastest non-magnetic models. However, even in the case of these fast models (dotted blue, cyan, and purple curves), the variation is less than an order of magnitude for central densities above 10$^4$ cm$^{-3}$. 

In the radial profiles, some variation in the abundance ratio occurs at large radii, but it is less than about an order of magnitude, at least in early times, and not much more than an order of magnitude in later times, even in the most-varying, fast non-magnetic models.

It is not surprising that the abundances of the two molecules are co-evolving, since the dominant reaction for the production of HCO$^+$ in our models involves CO: 
\begin{equation} 
{\rm CO  +  H_3^+  \rightarrow HCO^+ + H_2}\,,
\end{equation}
while the destruction of HCO$^+$ proceeds primarily through dissociative recombination with an electron:
\begin{equation}
{\rm HCO^+ + e^- \rightarrow CO + H}\,,
\end{equation}
which leads back to creating CO, forming a production/destruction loop for the two molecules.

\subsection{\rm{NO$^+$/NO}}

Figure \ref{ratioNOp} shows the same plots as before for the NO$^+$ / NO pair. This is the best example of co-evolution and chemical co-spatiality of an ion-neutral pair that we have identified in our coupled chemical/dynamical models. The variation of the central abundance ratio with central density in any single model is at most a factor of a few. These findings are also reflected in the radial profiles as well.

The reason for the co-evolution of the two molecules is that NO$^+$ is mainly produced through ionization of NO or charge-exchange reactions, also involving NO:
\begin{equation}
{\rm NO + H^+ \rightarrow NO^+ + H}
\end{equation} 
and 
\begin{equation}
{\rm NO + C^+ \rightarrow NO^+ + C}\,.
\end{equation} 

The ion is destroyed through dissociative recombination with an electron:
\begin{equation}
{\rm NO^+ +e^{-} \rightarrow O + N}\, ,
\end{equation} 
so the abundance of NO$^+$ is coupled to that of NO.
%%%%%%%%%%%%%%%%%%%%%%%%%%%%%%%%%%%%%%%%%%%%%%%%%%%%%%%%%%%%%%%%%%%%%%%%

\section{Discussion}\label{disc}

In addition to the co-spatiality of the three new ion-neutral pairs that we have identified, their observability and critical densities for excitation of their rotational transitions must also be considered to decide whether they are viable candidates for studies of the ambipolar ion-neutral drift in star-forming regions. 

The most frequently observed transitions of HCO$^+$ have  critical densities\footnote{All molecular data are taken from the Leiden Atomic and Molecular Database \citep{leiden05}, \tt{http://home.strw.leidenuniv.nl/$\sim$moldata/} } of the order of 10$^{5-6} {\rm \, cm^{-3}}$. NO generally has much lower critical densities, so overall unless higher critical density transitions of NO can be observed, this would pose a problem in the use of the HCO$^+$/NO pair. The reason is that NO would trace lower-density, higher-turbulence medium, which would lead to wider linewidths for NO, a trend that could mimic the effect observed in the HCO$^+$/HCN pair. Possible higher critical-density transitions would, for example, be the $J=7/2\rightarrow5/2$ transition of NO at $350.6$ GHz, with a critical density of $\simeq10^5 {\rm \, cm^{-3}}$. The multiplets of the first two rotational transitions of NO, i.e., $J=3/2\rightarrow1/2$ and $J=5/2\rightarrow3/2$ have been observed in both dark and massive clouds \citep{Gerin1992} at 150 GHz and 250 GHz with relative abundances calculated to vary from $10^{-7}-10^{-8}$. 

A similar problem arises with the HCO$^+$/CO pair, as CO, with generally low critical densities in usually observed transitions, traces more extended regions. Because the abundance of CO is high, such transitions would result in saturated and optically thick lines. To avoid this effect, one needs to use higher CO transitions. A possibility would be the $J=6\rightarrow5$ transition at $691$ GHz, with critical density around $10^5 {\rm \, cm^{-3}}$.

The NO$^+$/NO pair is in principle the best choice in terms of co-spatiality of the neutral and the ion, but although NO$^+$ has sub-millimeter and radio transitions identified in the laboratory \citep{Bowmanetal82}, the ion has not yet been observed in molecular clouds. This may be due to a general trend in which nitrogen-bearing molecules, with the exception of NH$_3$ and N$_2$H$^+$, have not been targeted for observations as much as the carbon-bearing species, in part because nitrogen chemistry in molecular clouds is even less understood (especially at low temperatures) than that of carbon (e.g., \citealp{HilyBlantetal2010}).

We note that our models do not include the effect of turbulence; as a result, the relative abundance variation that we find is an upper limit, since turbulent mixing would lessen this variation  \citep{Xieetal95}. Also, our models represent prestellar cores, and do not include the effect of feedback sources, because our parameter studies do not feature very large variations of the temperature, and UV ionization has not been accounted for. However, regions with feedback sources are avoided when applying the Houde et al. method.  

\acknowledgements{We thank Harold Yorke and Helmut Wiesemeyer for insightful and constructive comments that improved this paper.
T.H. is funded by the Alexander von Humboldt foundation. The project was supported in part by the NASA Origins of Solar Systems program. Part of this work was carried out at the Jet Propulsion Laboratory, California Institute of Technology under contract with the National Aeronautics and Space Administration.}

\end{document}